%% file: main.tex
\documentclass[prd,superscriptaddress,floatfix,nofootinbib,twocolumn]{revtex4-2}
\input{preamble}

\begin{abstract}
    The effect of black holes on entanglement harvesting has been of considerable interest over the past decade. Research involving stationary Unruh-DeWitt (UDW) detectors near a (2+1)-dimensional Bañados-Teitelboim-Zanelli (BTZ) black hole has uncovered phenomena such as entanglement shadows, entanglement amplification through black hole rotation, and differences between bipartite and tripartite entanglement. For a (1+1)-dimensional Schwarzschild black hole, it has been shown that two infalling UDW detectors can harvest entanglement from the scalar quantum vacuum even when separated by an event horizon. In this paper, we calculate the mutual information between two UDW detectors coupled to a massless quantum scalar field, with the detectors starting at rest and falling radially into a non-rotating (2+1)-dimensional BTZ black hole. The trajectory of the detectors includes regions where both detectors are switched on outside of the horizon; where one detector is switched on inside of the horizon while the other switches on outside; and where both detectors switch on inside of the horizon. We investigate different black hole masses, detector energy gaps, widths and temporal separations of the detector switching functions, and field boundary conditions. We find that black holes---even the simplest kind having constant curvature---significantly affect the correlation properties of quantum fields in the vacuum state. These correlations, both outside and inside the horizon, can be mapped out by infalling detectors.
\end{abstract}

\begin{document} 
\input{title_and_authors}

\maketitle
\flushbottom

\section{Introduction}\label{sec:introduction}

Although there exists no theory of quantum gravity, relativistic quantum information can help us investigate quantum effects in curved spacetime. In situations where particles are ill-defined, one approach is to use idealized particle detectors to give operational meaning to field excitations. The simplest such detector is the Unruh-DeWitt (UDW) detector~\cite{Unruh.effect, DeWitt1979}, a two-level quantum system that couples locally to a quantum field. Some well-known results obtained using this model are the Unruh effect~\cite{Unruh.effect,DeWitt1979,Davies1975} and entanglement degradation~\cite{Fuentes-Schuller:2004iaz}.  UDW detectors have been also used to study the Hawking effect~\cite{hawking_effect}, entanglement harvesting~\cite{Valentini:1991eah,EduardoDensityMatrix}, communication via quantum fields~\cite{Landulfo2016communication,Jonsson2018communication,Simidzija2020communication}, energy teleportation~\cite{Hotta2008energyteleportation}, and quantum Otto engines~\cite{Arias:2017kos,Gray:2018ifq,Gallock-Yoshimura:2023eaw}.  

With UDW detectors, one is interested in the transition probability of the detector from one state to the other, corresponding to the absorption of quanta from the field (or emission, in some cases). If there are two detectors, then one may also examine correlations between the detectors, providing information about the field correlation and entangling properties of spacetime.

Black hole spacetimes have been of particular interest. Static detectors outside of a Schwarzschild black hole have been found to respond thermally to the Hartle-Hawking state~\cite{Israel1976,HartleHawking1976,Hodgkinson:2014iua},  
and a stationary detector co-rotating with a (2+1)-dimensional Bañados-Teitelboim-Zanelli (BTZ) black hole was also found to thermalize at the Hawking temperature~\cite{hodgkinsonStaticStationaryInertial2012}. 
The transition rate, the time derivative of the transition probability, of a radially free-falling detector in the static BTZ case was also computed~\cite{hodgkinsonStaticStationaryInertial2012}  during the portion of the trajectory where the detector is outside of the horizon.

Further results of interest lie in two categories: those involving detectors falling across the horizon of a black hole
and those that investigate   correlations between detectors in the vicinity of a black hole. 
The transition rate of a detector freely falling across the horizon of a Bertotti-Robinson spacetime was found to decrease linearly 
with the infalling radius near the horizon
as the horizon is crossed~\cite{Conroy_2022}. The transition rate  for a detector falling into a (1+1)-dimensional Schwarzschild black hole~\cite{Juarez-Aubry:2014jba,JuarezPhDThesis} was found to experience a loss of thermal behaviour   during the infall. In another study~\cite{Juarez-Aubry:2021tae}, the transition rate of a detector falling into a (1+1)-dimensional Reissner-Nordstrom black hole was found to diverge at the Cauchy horizon for both the Unruh and Hartle-Hawking vacuum states.
Transition rates for detectors that are static outside of {geon black holes} have also been considered
\cite{Smith:2013zqa,Bhattacharya:2024jwm}. {$\mathbb{R}\text{P}^{2}$} geon spacetimes are  isometric to the static BTZ black hole outside the horizon but whose interiors are topologically non-trivial; the exterior Killing vector that generates exterior BTZ time translations does not extend to a Killing vector on the full spacetime. This nonstaticity bears an imprint on the response of the detector, providing a probe of otherwise hidden spacetime topology.

In higher dimensions, the response of a detector released from rest and falling freely across the
horizon of a (3+1)-dimensional Schwarzschild black hole
has been computed~\cite{Ng2022,ShallueCarroll2025}. The transition rate of an infalling detector in (2+1)-dimensional BTZ spacetime has also been computed for the static~\cite{MariaRosaBTZ},     rotating~\cite{Sijia2024}, and geon  \cite{Spadafora:2024gqj}
cases. In all of these scenarios,  new features known as ``glitches''---points at which the transition rate of the detector is non-differentiable---were discovered. Glitches occur near or inside the black hole and correspond to increased fluctuation and growth in the transition rate in those portions of the trajectory.

Entanglement harvesting---the process of using detectors to extract vacuum entanglement---has been investigated for static UDW detectors in the vicinity of a (2+1)-dimensional BTZ black hole~\cite{Henderson:2017yuv}. The black hole was found to inhibit entanglement harvesting, and to possess a ``shadow'' extending outside of the horizon in which no (bipartite) entanglement can be harvested. For a rotating BTZ black hole, angular momentum amplifies entanglement harvesting and diminishes the size of the shadow~\cite{Robbins:2020jca}. Another study~\cite{MutualInfoStatic2022} examined   the mutual information between static detectors near a (2+1)-dimensional BTZ black hole and found that, unlike  entanglement harvesting, mutual information does not vanish until one detector is placed at the horizon---i.e., there is no analogous shadow for mutual information. The case of three static detectors near a (2+1)-dimensional BTZ black hole has also been investigated~\cite{Tripartite2023}. It was found that one could harvest tripartite entanglement in regions where bipartite entanglement for static detectors would be forbidden. Several studies ~\cite{ali1,ali2,ali3,ali4} have investigated the transfer of correlations in a quantum field and the robustness of measures of entanglement between qubits when decoherence effects---such as Hawking radiation from a Schwarzschild black hole or the Gibbons-Hawking effect in a de Sitter vacuum---are present. Most recently, a study involving two static detectors in the vicinity of a Lorentz-violating (2+1)-dimensional BTZ black hole found that spontaneous Lorentz symmetry breaking suppresses entanglement harvesting while simultaneously enhancing mutual information harvesting~\cite{liu2025LorentzViolation}.

A study involving both infalling and correlation harvesting has been carried out  for two detectors in (1+1)-dimensional Schwarzschild spacetime~\cite{Gallock-Yoshimura:2021yok}.  Three scenarios were considered: in the first scenario, both detectors were static outside of the horizon; in the second scenario, one detector was static while the other fell freely into the black hole; and in the third scenario, both detectors, with some separation between them, fell freely into the black hole. In the latter scenarios, it was found that one could harvest entanglement from the vacuum even when the detectors were separated from one another by an event horizon. Moreover, the entanglement shadow was absent in the case of two free-falling detectors, since their relative gravitational redshift remained finite during horizon crossing.

In this paper, we 
carry out the first study of correlation harvesting for detectors freely falling into
a black hole that satisfies the Einstein equations. 
Specifically, we consider a non-rotating (2+1)-dimensional BTZ black hole and 
calculate the mutual information between two UDW detectors coupled to a massless quantum scalar field, with the detectors starting at rest and falling radially into and across the horizon. The trajectory of the detectors includes regions where both detectors are switched on outside of the horizon; where one detector is switched on inside of the horizon while the other switches on outside; and where both detectors switch on inside of the horizon. We investigate 
the dependence of mutual information on 
different black hole masses, detector energy gaps, widths and temporal separations of the detector switching functions, and field boundary conditions. 
The BTZ black hole simplifies calculations because the Wightman function of the conformally coupled scalar field can be computed as an image sum, which we then truncate after an appropriate number of terms. This procedure is considerably less tedious and complicated than the mode sum required for simulations a (3+1)-dimensional Schwarzschild black hole (see~\cite{Ng2022}). The dynamical problem of harvesting entanglement as detectors cross the event horizon is substantially more difficult; thus we examine mutual information (the sum of classical and quantum correlations) instead.

We find that mutual information is non-monotonic during detector infall: for a general trajectory, mutual information decreases initially, reaches an inflection point, and begins to increase as the detectors approach the singularity. This inflection point is formed by the growth of the correlation term outpacing the growth of the transition probabilities of the individual detectors during infall. In some cases, the inflection point is not attained before the singularity is reached; thus, mutual information decreases monotonically along the trajectory. In other cases, one only observes the monotonically increasing portion of the mutual information function. 
The main point is that infalling detectors can map out the correlation structure of the quantum vacuum via mutual information both outside and inside the horizon of the black hole.

The parameters with greatest impact on harvested mutual information are the mass of the black hole and those pertaining to the switching function. Larger black hole masses inhibit correlation harvesting, resulting in monotonically decreasing mutual information. As the mass becomes smaller, an inflection point appears. At small enough masses, the inflection point again vanishes, but with monotonically increasing mutual information. Thus, we observe that lighter black holes are statistically quieter. A similar effect is seen when the width of the switching function is decreased whilst holding other parameters constant. As the switching function becomes narrower, the mutual information transitions from monotonically decreasing to monotonically increasing, with an intermediary regime in which the mutual information possesses an inflection point.

The development of the inflection point, and how it evolves as a given parameter is varied, can be predicted by examining the relative positions of glitches in the transition probability and correlation terms of the detectors. As seen in previous studies involving freely-falling detectors in BTZ spacetime~\cite{MariaRosaBTZ,Sijia2024}, the Wightman function of the field can diverge at certain points along the detector trajectory. Employing a sharp switching function, the transition rate was found to be  non-differentiable (specifically, a cusp) when these divergent points coincide with the limits of integration. Glitches are a signature of BTZ spacetime, distinguishing it from pure anti-de Sitter (AdS) spacetime; consequently, in regions of the trajectory where one observes glitches, detector quantities, such as transition probability, will differ significantly from their values in pure AdS spacetime. In this study, we find that when glitches in the correlation term occur earlier than glitches in transition probability, correlation outpaces noise, and an inflection point forms. Note that we define a glitch as a point at which the Wightman function diverges at a limit of integration; we obtain the detector proper times at which these conditions are satisfied.

The outline of our  paper is as follows. In Section \ref{sec:setup}, we present the mathematical formalism required to compute mutual information between two detectors in (2+1)-dimensional BTZ spacetime. Section \ref{sec:results} contains the results of our simulations for different boundary conditions, black hole masses, switching function parameters, and detector energy gaps. In Section \ref{sec:conclusion}, we draw conclusions from our results and propose avenues of future work.

\section{Setup}\label{sec:setup}

We consider the Unruh-DeWitt model for two identical detectors coupled to a quantum massless scalar field in (2+1)-dimensional BTZ spacetime.

\subsection{BTZ Spacetime and the Quantum Vacuum}

The (2+1)-dimensional BTZ spacetime is a black hole spacetime of constant negative curvature. In the non-rotating case, the metric is
\begin{align}\label{eq:btz-metric}
    \dd s^2=-f(r)\dd t^2+\frac{\dd r^2}{f(r)}+r^2\dd \vp^2,
\end{align}
where $f(r)=\frac{r^2}{\ell^2}-M$. $M$ is the (dimensionless) mass of the black hole. Additionally, $t \in \R$, $r \in (0,\infty)$, and $\vp \in [0,2\pi)$. This metric (\ref{eq:btz-metric}) is a vacuum solution of Einstein's equations with cosmological constant $\Lambda=-1/\ell^2$, where $\ell(>0)$ is the anti-de Sitter (AdS) length. There is an event horizon at
\begin{align}\label{eq:rh}
    r_h=\ell\sqrt{M}.
\end{align}

By setting $M=1$ (without loss of generality) and $\vp\rightarrow y\in(-\infty,\infty)$ in (\ref{eq:btz-metric}), one obtains AdS$_3$-Rindler spacetime. Conversely, the BTZ spacetime can be obtained from AdS$_3$-Rindler by identifying $y\rightarrow\vp\in(0,\sqrt{M})$, and then rescaling $\vp$ and the radial and time coordinates to obtain (\ref{eq:btz-metric}); it is therefore locally equivalent to anti-de Sitter spacetime. We consider this fact when dealing with a quantum field in BTZ spacetime, where the correlation functions can be represented as a sum of correlators in AdS$_3$.

We consider a massless conformally coupled scalar field $\hat{\phi}(\sx)$ satisfying the Klein-Gordon equation
\begin{align}\label{eq:klein-gordon}
    (\Box-R/8)\hat{\phi}(\sx)=0,
\end{align}
where $\Box$ is the d'Alembert operator and $R$ is the Ricci scalar. The Wightman function $W_\mathrm{BTZ}(\sx,\sx'):=\langle 0|\hat{\phi}(\sx)\hat{\phi}(\sx')|0\rangle$ may be expressed as an image sum of Wightman functions in AdS spacetime. Specifically, the form is~\cite{LifschytzBTZ,carlipBTZ}
\begin{equation}\label{eq:wightman}
    \begin{split}
        & W_{\text{BTZ}}(\sx,\sx') = \sum_{n=-\infty}^\infty W_{\text{AdS}}(\sx,\Gamma^n\sx') \\
        & =\frac{1}{4\pi\sqrt{2}\ell} \sum_{n=-\infty}^\infty \left[\frac{1}{\sqrt{\sigma_\epsilon(\sx,\Gamma^n\sx')}}-\frac{\zeta}{\sqrt{\sigma_\epsilon(\sx,\Gamma^n\sx')+2}}\right],
    \end{split}
\end{equation}
where $\Gamma:(t,r,\vp)\mapsto(t,r,\vp+2\pi)$. The parameter $\zeta\in\{-1,0,1\}$ specifies the boundary conditions of the field at asymptotic infinity: Neumann, transparent, or Dirichlet respectively. If one selects Neumann or Dirichlet boundary conditions, the Wightman function corresponds to a Kubo-Martin-Schwinger (KMS) state that is analytic outside of the black hole horizon---i.e., the Hartle-Hawking state~\cite{Avis:1978,LifschytzBTZ}. There is no clear physical meaning for the state with transparent boundary condition~\cite{Avis:1978,LifschytzBTZ}. $\sigma_\epsilon(\sx,\Gamma^n\sx')$ is the squared geodesic separation (scaled by $\ell^2$) between two points in the covering AdS$_3$ spacetime. $\epsilon$ indicates that the Wightman function must be computed as a limit as $\epsilon\rightarrow 0_+$. $\sigma_\epsilon(\sx,\Gamma^n\sx')$ is given specifically by
\begin{align}
    \sigma_\epsilon(\sx,\Gamma^n\sx')&=\frac{rr'}{r_h^2}\cosh\left(\frac{r_h}{\ell}(\Delta\phi-2\pi n) \right)-1 \nonumber \\
    &- \frac{\sqrt{(r^2-r_h^2)(r'^2-r_h^2)}}{r_h^2}\cosh\left(\frac{r_h}{\ell^2}\Delta t-\ii \epsilon \right), \label{eq:sigma}
\end{align}
where $\Delta\phi=\phi-\phi'$ and $\Delta t=t-t'$.

\subsection{UDW Detectors and Mutual Information}

We consider a detector to be a pointlike qubit with states denoted by $|0\rangle$ and $|E\rangle$, for which the energy eigenvalues are $0$ and $E$ respectively. The detector moves on a timelike worldline $\sx(\tau)$, where $\tau$ is the proper time of the detector. Suppose we have two detectors, $A$ and $B$, and we let $\hat{\phi}$ be a massless scalar field {satisfying \eqref{eq:klein-gordon}}. The interaction Hamiltonian between a given detector and the field is
\begin{equation}\label{eq:hamiltonian}
    \hat{H}_{I, D}(\tau_D)=\lambda_D\chi_D(\tau_D)\hat{\mu}_D(\tau_D) \otimes \hat{\phi}(\sx_D(\tau_D))
\end{equation}
for $D \in \{A, B\}$, where $\hat{\mu}_D(\tau_D)=|E_D\rangle\langle 0|e^{\ii E_D\tau_D}+|0\rangle\langle E_D|e^{-\ii E_D\tau_D}$ is the detector's monopole moment operator  {in the interaction picture}, $\lambda_D$ is a coupling constant, and $\chi_D$ is the switching function, specifying how the detector is switched on and off. From this point on, we consider $\lambda_A=\lambda_B=\lambda$ and $E_A=E_B=E$.

The total interaction Hamiltonian with respect to BTZ coordinate time $t$ is
\begin{align}\label{eq:total-hamiltonian}
    \hat{H}_I(t)=\frac{\dd \tau_A}{\dd t}\hat{H}_{I, A}(\tau_A(t))+\frac{\dd \tau_B}{\dd t}\hat{H}_{I, B}(\tau_B(t)),
\end{align}
and the time evolution operator is
\begin{align}\label{eq:u-operator}
    \hat{U}_I=\mathcal{T} \exp{\left( -\ii\int_\R\dd t~\hat{H}_I(t) \right)}.
\end{align}

For small $\lambda$, the time evolution operator may be expanded as a Dyson series:
\begin{align}\label{eq:dyson}
    \hat{U}_I=\mathbbm{1}+\hat{U}^{(1)}+\hat{U}^{(2)}+\mathcal{O}(\lambda^3),
\end{align}
where
\begin{align}
    &\hat{U}^{(1)}=-\ii \int_{-\infty}^\infty \dd t~\hat{H}_I(t), \label{eq:dyson1} \\ 
    &\hat{U}^{(2)}=-\int_{-\infty}^\infty \dd t_1 \int_{-\infty}^{t_1} \dd t_2 ~\hat{H}_I(t_1)\hat{H}_I(t_2), \label{eq:dyson2}
\end{align}
and so on. If we assume that the detectors are initially in their ground state and uncorrelated, then the final density matrix, after tracing out the field, is~\cite{EduardoDensityMatrix}
\begin{align}\label{eq:density-matrix}
    \rho_{AB}=\begin{pmatrix}
        1-P_A-P_B & 0 & 0 & X \\
        0 & P_B & C & 0 \\
        0 & C^* & P_A & 0 \\
        X^* & 0 & 0 & 0
    \end{pmatrix}+\mathcal{O}(\lambda^4),
\end{align}
where
\begin{widetext}
\begin{align}
    &P_D=\lambda^2\int \dd\tau_D \dd\tau_D' ~\chi_D(\tau_D)\chi_D(\tau_D')e^{-\ii E(\tau_D-\tau_D')}W_{\mathrm{BTZ}}(\sx_D(\tau_D),\sx_D(\tau_D')) \quad \mathrm{for} \quad D \in \{A, B\}, \label{eq:p} \\
    &C=\lambda^2\int \dd\tau_A \dd\tau_B ~\chi_A(\tau_A)\chi_B(\tau_B)e^{-\ii E(\tau_A-\tau_B)}W_{\mathrm{BTZ}}(\sx_A(\tau_A),\sx_B(\tau_B)), \label{eq:c} \\
    &X=-\lambda^2\int \dd\tau_A \dd\tau_B ~\chi_A(\tau_A)\chi_B(\tau_B)e^{-\ii E(\tau_A+\tau_B)}\left[\theta(t_B-t_A) W(\sx_A(\tau_A),\sx_B(\tau_B))+ \theta(t_A-t_B)W(\sx_B(\tau_B),\sx_A(\tau_A)) \right]. \label{eq:x}
\end{align}
\end{widetext}

The reduced density matrices of the individual detectors are
\begin{align}\label{eq:reduced-density}
    \rho_D=\begin{pmatrix}
        1-P_D & 0 \\
        0 & P_D
    \end{pmatrix}+\mathcal{O}(\lambda^4)
\end{align}
for $D \in \{A, B\}$. Thus, $P_A$ and $P_B$ are the transition probabilities of detectors $A$ and $B$ respectively. $C$ (henceforth referred to as the correlation term) is used to compute mutual information, while $X$ is used to compute measures of entanglement  {between the detectors}.

The mutual information between the two detectors is \cite{Nielsen:2012yss} 
\begin{align}\label{eq:mutual-info}
    \mathcal{I}_{AB} =~&\mathcal{L}_+ \ln \mathcal{L}_+ + \mathcal{L}_- \ln \mathcal{L}_- \nonumber \\
    &- P_A \ln P_A - P_B \ln P_B + \mathcal{O}(\lambda^4)
\end{align}
where
\begin{align}\label{eq:l-pm}
    \mathcal{L}_\pm=\frac{1}{2}\left(P_A+P_B \pm \sqrt{(P_A-P_B)^2+4|C|^2} \right).
\end{align}

Note that when $C=0$, $\mathcal{I}_{AB}=0$ as well. Non-perturbatively, $\rho_{ii}\rho_{jj}\geq |\rho_{ij}|^2$. Neglecting $\mathcal{O}(\lambda^4)$ contributions, this implies that $P_A P_B \geq |C|^2$.

\subsection{Trajectories and Computations}

The trajectory taken by the detectors is~\cite{hodgkinsonStaticStationaryInertial2012}
\begin{align}
    &t_D(\tau_D)=\frac{r_h}{M} \arctanh{\left(\frac{\tan{(\tau_D/\ell)}}{\sqrt{q^2-1}} \right)}, \label{eq:traj-t} \\
    &r_D(\tau_D)=qr_h \cos{(\tau_D/\ell)}, \label{eq:traj-r} \\
    &\phi_D(\tau_D)=\phi_0 \label{eq:traj-phi}
\end{align}
for $D \in \{A, B\}$, where $q>1$ is a dimensionless factor controlling the initial distance (at $t_D=\tau_D=0$) 
of the detectors from the centre of the black hole. We employ switching functions of 
compact support  
\begin{align}
    &\chi_A(\tau_A)=\begin{cases}
        \cos^4\left(\frac{\pi(\tau_A-\tau_{\textrm{mid}})}{\Delta\tau} \right), & \text{if } |\tau_A-\tau_{\textrm{mid}}|<\frac{\Delta\tau}{2} \\
        0, & \text{otherwise}
    \end{cases}, \label{eq:switching-a} \\
    &\chi_B(\tau_B)=\begin{cases}
        \cos^4\left(\frac{\pi(\tau_B+T-\tau_{\textrm{mid}})}{\Delta\tau} \right), & \text{if } |\tau_B+T-\tau_{\textrm{mid}}|<\frac{\Delta\tau}{2} \\
        0, & \text{otherwise}
    \end{cases}, \label{eq:switching-b}
\end{align}
where $\Delta\tau$ is the width of the support of the switching function and $\tau_{\textrm{mid}}$ is the location of its peak (for detector A). 
We select $T>\Delta\tau$ so that the switching functions of the two detectors do not overlap. Thus we have two UDW detectors that fall from rest at the same initial radial position but switch on at different times, with B switching on first. For future calculations, we define $\tau$ and $\tau_0$ such that $\Delta\tau=\tau-\tau_0$ and $\tau_{\mathrm{mid}}=\frac{1}{2}(\tau+\tau_0)$. We also rescale $\tau\rightarrow\tau/\ell$, $\Delta\tau\rightarrow\Delta\tau/\ell$, $E\to E\ell$, and so on, for convenience.

With this trajectory, the transition probability of detector $A$, given in Eq. (\ref{eq:p}), becomes~\cite{hodgkinsonStaticStationaryInertial2012}
\begin{widetext}
\begin{align}
    P_A=\frac{1}{4}\int_{\tau_0}^\tau \dd u ~\left[\chi_A(u)\right]^2+2\int_{\tau_0}^\tau \dd u ~\chi_A(u) \int_0^{u-\tau_0} \dd s ~\chi_A(u-s)\mathrm{Re}\left[e^{-\ii Es}W(u,u-s) \right], \label{eq:p-sim}
\end{align}
and the correlation term in Eq. (\ref{eq:c}) becomes
\begin{align}
    C=\int_{\tau_0}^\tau \dd u ~\chi_A(u) \int_0^{u-\tau_0} \dd s ~\chi_A(u-s)e^{-\ii ET}\left(e^{-\ii Es}W(u,u-s-
    T)+e^{\ii Es}W(u-s,u-T) \right), \label{eq:c-sim}
\end{align}
\end{widetext}
where
\begin{align}
    &W(u,v)=\frac{1}{4\pi\sqrt{2}}\sum_{n=-\infty}^\infty \left(\frac{1}{\sqrt{\Delta X_n^2}}
    -\frac{\zeta}{\sqrt{\Delta X_n^2+2}} \right), \label{eq:wightman2} \\
    &\Delta X_n^2(u,v)=-1+K_n\cos(u)\cos(v)+\sin(u)\sin(v), \label{eq:geo-dist} \\
    &K_n=1+2q^2\sinh^2\left(\sqrt{M}n\pi \right). \label{eq:kn}
\end{align}

We write $\Delta X_n^2$, rather than $\sigma_\epsilon(\sx,\Gamma^n\sx')$, for the squared geodesic distance to distinguish that there is no $\epsilon$-regulator in the Wightman function here. To obtain Eq. (\ref{eq:p-sim}), we have evaluted Eq. (\ref{eq:p}) in the limit as $\epsilon\rightarrow0_+$; this limit has a relatively simple analytic form. We also evaluate the limit as $\epsilon\rightarrow0_+$ to go from Eq. (\ref{eq:c}) to (\ref{eq:c-sim}), but this limit is straightforward since we have imposed $T>\Delta\tau$.

Several comments should be made at this point. First, the transition probability of detector $B$ is the same as that of detector $A$ but shifted by $T$ to the right. That is, $P_B(\tau)=P_A(\tau-T)$. Second, there are two instances of $\chi_A$ in Eq. (\ref{eq:c-sim}) because a change of variables that occurs when simplifying $C$ turns $\chi_B$ into $\chi_A$. Third, in Eq. (\ref{eq:p-sim}) and (\ref{eq:c-sim}), we express $P_A$ and $C$ in units of the dimensionless quantity $\lambda^2\ell$, which we set equal to $1$ when presenting our results. However, if $P_D \rightarrow \lambda^2\ell P_D$ for $D \in \{A, B\}$ and $C \rightarrow \lambda^2\ell C$, one can show using Eq. (\ref{eq:mutual-info}) and (\ref{eq:l-pm}) that $\mathcal{I}_{AB} \rightarrow \lambda^2\ell \mathcal{I}_{AB}$. Thus, $P_A$, $P_B$, $C$, and $\mathcal{I}_{AB}$ possess a degree of freedom that is a straightforward scaling.

We may decompose the transition probability and correlation term as
\begin{align}
    P_D&=P_D^{n=0}+P_D^{n\neq 0} \label{eq:p-decomp} \\
    C&=C^{n=0}+C^{n\neq 0} \label{eq:c-decomp}
\end{align}
for $D \in \{A, B\}$, where the $n=0$ term gives the transition probability and correlation term respectively for the detectors in AdS$_3$ spacetime, while the $n\neq 0$ terms arise only in BTZ spacetime. Furthermore, we can determine, by inserting Eq. (\ref{eq:wightman2}-\ref{eq:kn}) into Eq. (\ref{eq:p-sim}) and Eq. (\ref{eq:c-sim}), that both $P_D$ and $C$ are invariant under $n\rightarrow -n$. Thus, one only needs to compute the positive terms in the image sum and multiply by 2.

In Eq.~(\ref{eq:wightman2}), it is possible for $\Delta X_n^2$ to vanish within the range of integration. Since these singularities are of the inverse square root type, they are integrable. When a singularity of the Wightman function coincides with one of the bounds of integration in Eq. (\ref{eq:p-sim}), the corresponding point in the trajectory is called a ``glitch.'' This terminology comes from previous work calculating the transition rate of detectors in (2+1)-dimensional BTZ spacetime~\cite{MariaRosaBTZ,Sijia2024}, where glitches could be observed as non-differentiable points.

The equation $a\sin x+b \cos x-c=0$ has solutions
\begin{align} \label{eq:s12}
    s_{1,2}(a,b,c)=\mathrm{arctan2}&\left(bc\mp a\sqrt{a^2+b^2-c^2}, \right.\\
    &\left. ~ac\pm b\sqrt{a^2+b^2-c^2} \right) \nonumber,
\end{align}
where $\mathrm{arctan2}(x,y)$ is the argument of the complex number $x+\ii y$. Let us first consider the transition probability. If we set $v=u-s$ in Eq. (\ref{eq:geo-dist}) and take $x=u-s$, then the equation has roots at
\begin{align} \label{eq:s-roots}
    s^*=u^*-s_{1,2}(\sin u, K_n\cos u, 1).
\end{align}

We wish to find pairs $(s^*,u^*)$ such that both $s^*$ and $u^*$ are equal to one of the limits of their respective integrals in Eq. (\ref{eq:p-sim}). This condition is not always possible to satisfy over the domain of integration for all combinations of upper and lower limits. For example, if we choose $s^*=0$ (the lower bound of the $s$-integral), then Eq. (\ref{eq:s-roots}) can only be satisfied by $u^*=\frac{\pi}{2}$, but then we cannot simultaneously satisfy $u^*=\tau_0$ (the lower bound of the $u$-integral), since that would imply a detector switching on at $\tau=\pi/2$ (i.e., the black hole singularity). It turns out that for $P_A$, there is only one set of glitches occurring before the singularity---namely, those at
\begin{align}
    \tau_A^*=\frac{1}{2}s_{1,2}(&(K_n-1)\sin \Delta\tau,(K_n-1)\cos \Delta\tau, \\
    &2-(K_n+1)\cos \Delta\tau), \nonumber
\end{align}
where we have substituted $\Delta\tau=\tau-\tau_0$. While the notation $s_{1,2}$ implies two solutions, one of the solutions lies outside of the domain of integration, and so is discarded.  It also happens that there are no glitches originating from the term in Eq. (\ref{eq:p-sim}) that is proportional to $\zeta$. This statement is true in non-rotating BTZ spacetime, but not in a general, rotating BTZ spacetime~\cite{Sijia2024}. The glitches in $P_B$ are identical to those in $P_A$, but shifted by $T$ to the right. That is, $\tau_B^*=\tau_A^*+T$.

The geodesic separation $\Delta X_n^2$ vanishes when two points are separated by a null ray. This is impossible when $n=0$, since any two points on the trajectory of the same detector must be timelike separated. However, with the cylindrical identification $\vp\rightarrow \vp+2\pi$ in BTZ spacetime, null rays can wrap around the $\vp$ direction to intersect with another point in the same trajectory. This results in an infinite number of glitches indexed by $n$. For the transition probability $P_A$, glitches correspond to switch-off events $\sx(\tau)$ that are null separated and in the future of identified switch-on events $\Gamma^n \sx(\tau_0)$, where $\Gamma:(t,r,\vp)\mapsto(t,r,\vp+2\pi)$. Similarly, for $P_B$, glitches correspond to switch-off events $\sx(\tau-T)$ that are null separated and in the future of identified switch-on events $\Gamma^n \sx(\tau_0-T)$.

The glitches in the correlation term are more complicated. In Eq. (\ref{eq:c-sim}), $C$ has two terms: one term proportional to $\mathrm{e}^{-\ii Es}$ (call it the \textit{positive} term) and another proportional to $\mathrm{e}^{\ii Es}$ (the \textit{negative} term). Since the arguments in the Wightman function differ in the positive and negative terms, the two cases will form different glitches. As before, however, there are no glitches arising from the component of the Wightman function that is proportional to $\zeta$.

    For each term, positive and negative, there are three types of glitches that can come into play. Type I glitches correspond to switch-on events $\sx(\tau_0)$ of detector $A$ that are null separated from identified switch-on events $\Gamma^n \sx(\tau_0-T)$ of detector $B$; type II glitches correspond to switch-off events $\sx(\tau)$ of detector $A$ that are null separated from identified switch-off events $\Gamma^n \sx(\tau-T)$ of detector $B$; and type III glitches correspond to null separation between a switch-on and a switch-off event, either switch-off events $\sx(\tau)$ of detector $A$ that are null separated from identified switch-on events $\Gamma^n \sx(\tau_0-T)$ of detector $B$, or switch-on events $\sx(\tau_0)$ of detector $A$ that are null separated from identified switch-off events $\Gamma^n \sx(\tau-T)$ of detector $B$.
    

\begin{widetext}
For the positive term, the locations of the three types of glitches are given by
\begin{align}
    &\tau_{\mathrm{I},+}^*=\Delta\tau+\frac{1}{2}s_{1,2}((K_n-1)\sin T,(K_n-1)\cos T,2-(K_n+1)\cos T), \label{eq:tau11} \\
    &\tau_{\mathrm{II},+}^*=\frac{1}{2}s_{1,2}((K_n-1)\sin T,(K_n-1)\cos T,2-(K_n+1)\cos T), \label{eq:tau12} \\
    &\tau_{\mathrm{III},+}^*=\frac{1}{2}s_{1,2}((K_n-1)\sin(T+\Delta\tau),(K_n-1)\cos(T+\Delta\tau),2-(K_n+1)\cos(T+\Delta\tau)). \label{eq:tau13}
\end{align}

For the negative term, the locations of the three types of glitches are given by
\begin{align}
    &\tau_{\mathrm{I},-}^*=\Delta\tau+\frac{1}{2}s_{1,2}((L_n+1)\sin T,(L_n+1)\cos T,2-(L_n-1)\cos T), \label{eq:tau21} \\
    &\tau_{\mathrm{II},-}^*=\frac{1}{2}s_{1,2}((L_n+1)\sin T,(L_n+1)\cos T,2-(L_n-1)\cos T), \label{eq:tau22} \\
    &\tau_{\mathrm{III},-}^*=\Delta\tau+\frac{1}{2}s_{1,2}((L_n+1)\sin(T-\Delta\tau),(L_n+1)\cos(T-\Delta\tau),2-(L_n-1)\cos(T-\Delta\tau)), \label{eq:tau23}
\end{align}
\end{widetext}
where
\begin{align}
    L_n=-1+2q^2\cosh^2\left(\sqrt{M}n\pi \right).
\end{align}

Note that while each type of glitch for both terms in $C$ \textit{can} occur along the detectors' trajectory (i.e., before the detectors reach singularity), they do not necessarily occur before reaching the singularity. Any glitches that occur at $\tau>\frac{\pi}{2}$ are unphysical.

\section{Results}\label{sec:results}

In this section, we present our results for mutual information between two infalling detectors for different field boundary conditions, black hole masses, switching function parameters, and detector energy gaps. We combine these results with observations of glitch behaviour to generate a comprehensive picture of each trend. In all of our results, the initial position of both detectors is $r_0=5r_h$. Each detector takes proper time $\tau/\ell=\pi/2$ from when it is released to reach the singularity. If $\tau$ is the upper bound of the support of $\chi_A$ (i.e., the later switching function), then we start the simulation at time $\tau=T+\Delta\tau$ so that the earlier switching function, $\chi_B$, has support only when both detectors are already falling.

In our computations, we truncate the image sum at a term where the absolute value of the next term's  contribution to $P_A$, $P_B$, and $C$ is less than $10^{-12}$ over the whole trajectory. We find that the mutual information is also accurate to $10^{-12}$ if we impose these precision requirements, for the investigated parameter space.

\subsection{Boundary Conditions} \label{subsec:results-a}

\begin{figure}
    \centering
    \includegraphics[width=\linewidth]{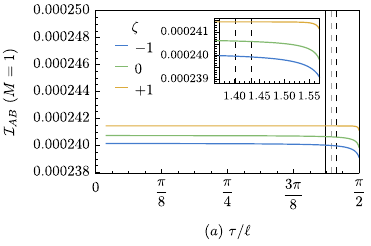}
    \includegraphics[width=\linewidth]{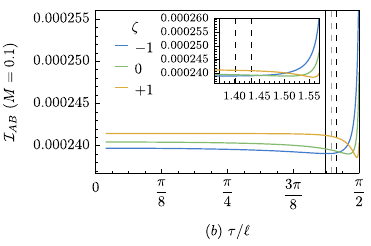}
    \includegraphics[width=\linewidth]{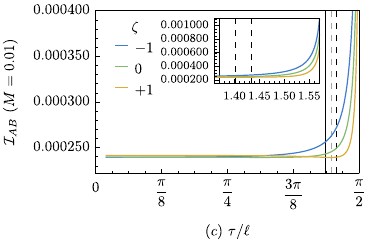}
    \caption{The mutual information $\mathcal{I}_{AB}$ between the two UDW detectors, with $r_0/r_h=5$, $E\ell=1$, $\Delta\tau/\ell=\pi/100$, and $T/\ell=\Delta\tau/\ell+0.001$, for (a) $M = 1$, (b) $M = 0.1$, and (c) $M = 0.01$. We calculate the image sum from $n=-N$ to $n=+N$ for $N=8$, $22$, and $70$ respectively. The solid vertical line indicates when the support of $\chi_A$ reaches the horizon of the black hole. The dashed vertical lines indicate when the support of $\chi_B$ reaches the horizon (left line) and when the supports of both switching functions are fully contained inside the horizon (right line). The right edge of the plot is the time to singularity.
    }
    \label{fig:boundaries}
\end{figure}

Fig. \ref{fig:boundaries} shows the mutual information $\mathcal{I}_{AB}$ between the two detectors as a function of the proper time of detector $A$ for the three boundary conditions, $\zeta=-1$, $0$, and $1$, given black hole mass of (a) $M=1$, (b) $M=0.1$, and (c) $M=0.01$. In all three plots, we fix $E\ell=1$, $\Delta\tau/\ell=\pi/100$, and $T/\ell=\Delta\tau/\ell+0.001$. We select these masses only to provide some representative plots to compare the boundary conditions. A detailed treatment of the effect of varying mass is given in the next section.

In general, mutual information is near constant when the detectors are far from the horizon. This observation is consistent with the fact that BTZ spacetime is asymptotically AdS;  the mutual information between two ``infalling'' detectors in AdS$_3$ is constant, since $P_D$ and $C$ are constant. As the detectors approach the black hole, the mutual information varies, usually exhibiting a dip. In some cases, this dip continues to the singularity, as in Fig. \ref{fig:boundaries}(a). In other cases, the mutual information forms an inflection point, reaching a minimum and then beginning to increase as the singularity is approached (Fig. \ref{fig:boundaries}(b) and \ref{fig:boundaries}(c)). This increase becomes more dramatic as black hole mass decreases.

For all three boundary conditions, mutual information exhibits the same changes when mass (or any other parameter in the problem) is varied. The only difference between the $\zeta$'s is the point in the parameter space at which different phenomena emerge. For example, in Fig. \ref{fig:boundaries}(c), the mutual information when $\zeta=-1$ is monotonically increasing, whereas  the mutual information when $\zeta=0$ and $\zeta=1$ still possesses an inflection point. Although not shown in Fig. \ref{fig:boundaries}, the inflection point for $\zeta=-1$ also forms at larger values of $M$ than for $\zeta=0$ and $\zeta=1$. In general, $\zeta=-1$ is the most \textit{sensitive} case, in that, if an inflection point were to form from gradually increasing or decreasing a certain parameter, it would form first in the $\zeta=-1$ case. The $\zeta=0$ case is the second most sensitive, while the $\zeta=1$ case is the least sensitive. In the remainder of this paper, we fix $\zeta=-1$, because it is computationally easier to probe the relevant parameter space for this case, and also because the $\zeta=-1$ case can be readily interpreted as a Hartle-Hawking state.

\subsection{Varying Black Hole Mass} \label{subsec:results-b}

\begin{figure*}[t]
    \centering
    \includegraphics[width=0.49\linewidth]{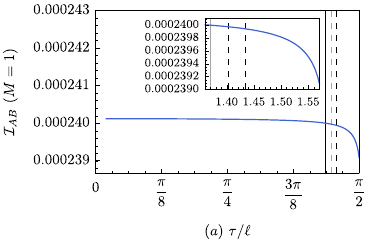}
    \includegraphics[width=0.49\linewidth]{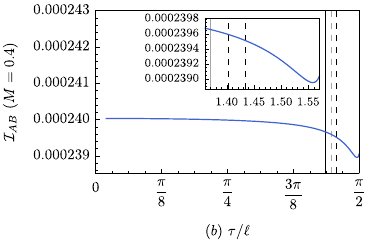}
    \includegraphics[width=0.49\linewidth]{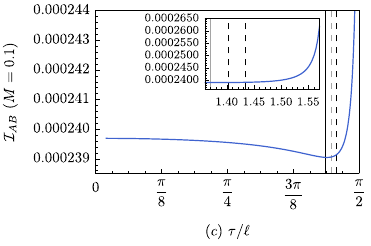}
    \includegraphics[width=0.49\linewidth]{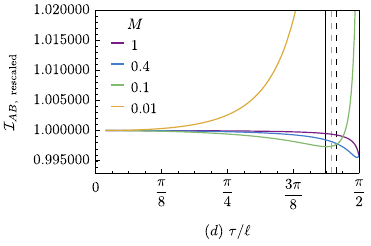}
    \caption{Plots (a), (b), and (c) show the mutual information $\mathcal{I}_{AB}$ between both detectors for $M=1$, $0.4$, and $0.1$ respectively. Observe the formation of the inflection point. Plot (d) gives the rescaled mutual information for $M=1$, $0.4$, $0.1$, and $0.01$, such that each function begins at $1$. For all plots, we fix $r_0/r_h=5$, $E\ell=1$, $\zeta=-1$, $\Delta\tau/\ell=\pi/100$, and $T/\ell=\Delta\tau/\ell+0.001$. For the masses in decreasing order, we calculate the image sum from $n=-N$ to $n=+N$ for $N=8$, $12$, $22$, and $70$ respectively. The solid, vertical line indicates when the support of $\chi_A$ reaches the horizon of the black hole. The dashed, vertical lines indicate when the support of $\chi_B$ reaches the horizon (left line) and when the supports of both switching functions are fully contained inside the horizon (right line). The right edge of the plot is the time to singularity. In (d), the truncated green curve ($M=0.1$) rises to $1.092549$ just before $\tau/\ell=\pi/2$ while the truncated yellow curve ($M=0.01$) rises to $4.562299$.
    }
    \label{fig:masses1}
\end{figure*}

Fig. \ref{fig:masses1}(a)-(c) show the mutual information $\mathcal{I}_{AB}$ between the two detectors as a function of the proper time of detector $A$, given black hole masses $M=1$, $0.4$, and $0.1$ respectively. In each plot, we fix $r_0/r_h=5$, $E\ell=1$, $\zeta=-1$, $\Delta\tau/\ell=\pi/100$, and $T/\ell=\Delta\tau/\ell+0.001$. In Fig. \ref{fig:masses1}(d), we compare the mutual information for $M=1$, $0.4$, $0.1$, and $0.01$ on the same plot. The comparison is facilitated by rescaling $\mathcal{I}_{AB}$ such that the first point on each curve takes a value of 1.

From Fig. \ref{fig:masses1}(a)-(c), we observe the  formation of an inflection point in $\mathcal{I}_{AB}$ when the mass is decreased to roughly $M=0.4$. This inflection point first appears near the singularity (i.e., the right edge of the plot) and gradually migrates to earlier times in the trajectory. To the right of the inflection point, the mutual information is monotonically increasing. From Fig. \ref{fig:masses1}(d), we observe that the growth of mutual information past the inflection point is rapid: the magnitude of the increase in $\mathcal{I}_{AB}$ is much larger than the magnitude of the decrease leading up to the inflection point. The minimum mutual information attained as a fraction of the mutual information at the first observed point tends to $1$ as $M$ decreases.

We emphasize that, as the mass of the black hole decreases, the mutual information that is harvested by the detectors increases. This increase corresponds to the formation of an inflection point, beyond which mutual information grows rapidly. In the case of $M=0.01$ in Fig. \ref{fig:masses1}(d), the inflection point is so far to the left that it no longer exists, and $\mathcal{I}_{AB}$ increases dramatically from early times.

\begin{figure*}[t]
    \centering
    \includegraphics[width=0.328\linewidth]{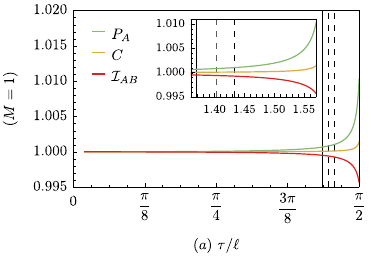}
    \includegraphics[width=0.328\linewidth]{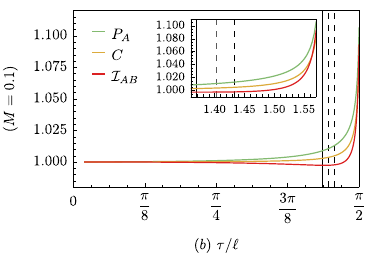}
    \includegraphics[width=0.328\linewidth]{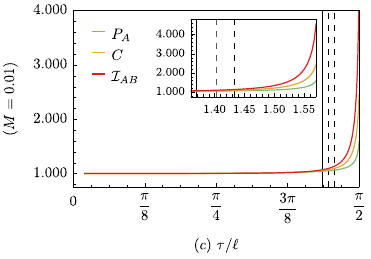}
    \caption{Comparison of transition probability $P_A$, correlation term $C$, and mutual information $\mathcal{I}_{AB}$ for (a) $M=1$, (b) $M=0.1$, and (c) $M=0.01$. $P_A$, $C$, and $\mathcal{I}_{AB}$ are rescaled such that each function begins at $1$. Observe that $C$ grows faster relative to $P_A$ as mass decreases, driving the growth of $\mathcal{I}_{AB}$. We fix $r_0/r_h=5$, $E\ell=1$, $\zeta=-1$, $\Delta\tau/\ell=\pi/100$, and $T/\ell=\Delta\tau/\ell+0.001$, and calculate the image sum from $n=-N$ to $n=+N$ for $N=8$, $22$, and $70$ respectively. The solid, vertical line indicates when the support of $\chi_A$ reaches the horizon of the black hole. The dashed, vertical lines indicate when the support of $\chi_B$ reaches the horizon (left line) and when the supports of both switching functions are fully contained inside the horizon (right line). The right edge of the plot is the time to singularity.
    }
    \label{fig:masses2}
\end{figure*}

In Fig. \ref{fig:masses2}, we compare the transition probability, correlation term, and overall mutual information for black hole masses (a) $M=1$, (b) $M=0.1$, and (c) $M=0.01$. In each plot, we fix $r_0/r_h=5$, $E\ell=1$, $\zeta=-1$, $\Delta\tau/\ell=\pi/100$, and $T/\ell=\Delta\tau/\ell+0.001$. Note that we only show $P_A$, the transition probability of detector $A$, since $P_B(\tau)=P_A(\tau-T)$ (i.e., $P_B$ is $P_A$ with a temporal shift). We rescale $P_A$, $C$, and $\mathcal{I}_{AB}$ so that they all begin at $1$.

In Fig. \ref{fig:masses2}(a), for the large mass case, $P_A$ grows more quickly than $C$ as the detectors approach the singularity. As a result, mutual information decreases monotonically. In Fig. \ref{fig:masses2}(b), for a smaller mass, $P_A$ also grows more quickly than $C$, but the difference is not as pronounced. $C$ grows quickly enough that $\mathcal{I}_{AB}$ achieves an inflection point and increases towards the singularity. Finally, in Fig. \ref{fig:masses2}(c), for the smallest black hole mass, $C$ grows more quickly than $P_A$. $\mathcal{I}_{AB}$ also grows dramatically as a result. If one interprets $P_A$ and $P_B$ as noise and $C$ as correlation, then we observe that black holes with smaller masses are statistically quieter.

\begin{figure*}[t]
    \centering
    \includegraphics[width=0.32\linewidth]{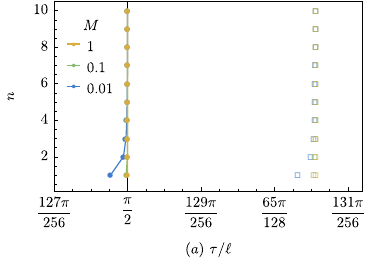}
    \includegraphics[width=0.3154\linewidth]{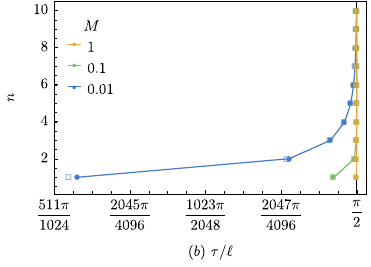}
    \includegraphics[width=0.318\linewidth]{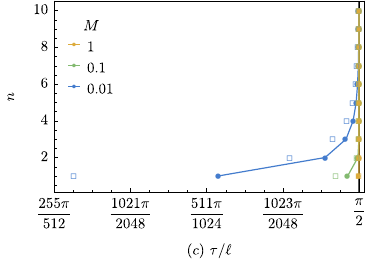}
    \includegraphics[width=0.32\linewidth]{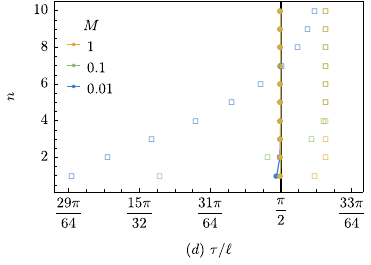}
    \includegraphics[width=0.317\linewidth]{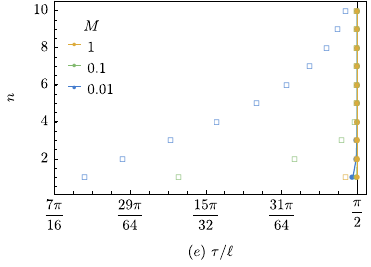}
    \includegraphics[width=0.319\linewidth]{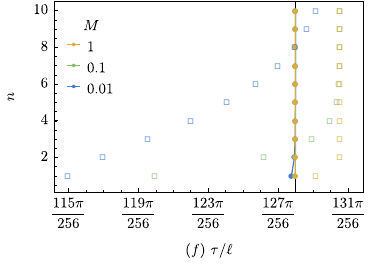}
    \caption{Comparison of the glitches in $C$ (unfilled squares) to the glitches in $P_A$ (filled circles) for $M=1$, $0.1$, and $0.01$. Points having the same colour correspond to the same mass. Plots (a), (b), and (c) show the type I, II, and III glitches respectively from the positive term in $C$, while plots (d), (e), and (f) show the type I, II, and III glitches respectively from the negative term in $C$. The $P_A$ glitches are the same in all plots. We fix $r_0/r_h=5$, $\Delta\tau/\ell=\pi/100$, and $T/\ell=\Delta\tau/\ell+0.001$. The vertical axis indicates the term number $n$ in the image sum from which the glitch arises. The solid vertical line indicates when the black hole singularity is reached.
    }
    \label{fig:masses3}
\end{figure*}

Fig. \ref{fig:masses3} shows the glitches of the correlation term $C$ relative to the transition probability $P_A$ for $M=1$, $0.1$, and $0.01$. In each plot, we fix $r_0/r_h=5$, $\Delta\tau/\ell=\pi/100$, and $T/\ell=\Delta\tau/\ell+0.001$. Note again that we only include $P_A$, the transition probability of detector $A$, since $P_B(\tau)=P_A(\tau-T)$. We extend the plot past the time to singularity $\tau/\ell=\pi/2$ in order to show the structure of the glitches.

In all cases, we observe that the glitches of $P_A$ and $C$ move to the left as the black hole mass decreases. However, the glitches of $C$ move leftward more quickly than those of $P_A$, when varying $M$. The occurrence of glitches is \textit{correlated} with the appearance of non-AdS features in $P_A$ and $C$---namely, the monotonic growth of both functions as the detectors approach the black hole. In pure AdS$_3$, we would expect constant functions. We stress the idea of \textit{correlation}, because one cannot say that, if the glitches of $C$ occur before those of $P_A$, there will definitely be an inflection point in $\mathcal{I}_{AB}$, or vice versa. In fact, as we saw in Fig. \ref{fig:boundaries}, the presence or absence of an inflection point for a given set of parameters depends on the boundary condition of the field, whereas the locations of the glitches in $P_A$ and $C$ do not.

However, the glitches do predict the overall trend of mutual information when varying a given parameter. In the case of black hole mass, we observe that the rapid leftward expansion of the glitches in $C$ relative to their counterparts in $P_A$ as $M$ decreases predicts the phenomenon in Fig. \ref{fig:masses2} (that is, that $C$ grows more rapidly than $P_A$), which in turn predicts the formation of the inflection point in $\mathcal{I}_{AB}$.

Since $C$ gives rise to six types of glitches, one may wish to consider some kind of ``average'' behaviour among the six sets of glitches. For example, in Fig. \ref{fig:masses3}(d), the $n=1$ and $n=2$ glitches in $C$ when $M=0.1$ or $M=0.01$ clearly occur before the $n=1$ and $n=2$ glitches in $P_A$, whereas in Fig. \ref{fig:masses3}(a), all of the glitches in $C$ occur after those in $P_A$ (and, in fact, the glitches in $C$ are unphysical). However, we find that it is unnecessary to average over the glitches: as long as even one set of glitches in $C$ moves leftward relative to the glitches in $P_A$, we can expect an inflection point to form.

\subsection{Varying the Width of the Switching Function}  \label{subsec:results-c}

\begin{figure*}[t]
    \centering
    \includegraphics[width=0.49\linewidth]{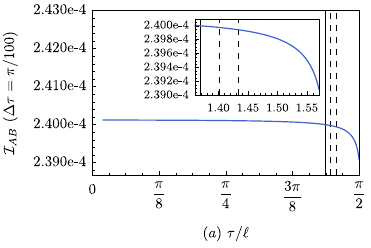}
    \includegraphics[width=0.49\linewidth]{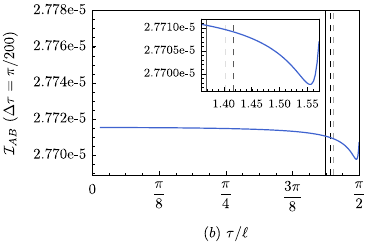}
    \includegraphics[width=0.49\linewidth]{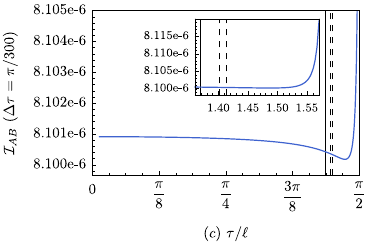}
    \includegraphics[width=0.49\linewidth]{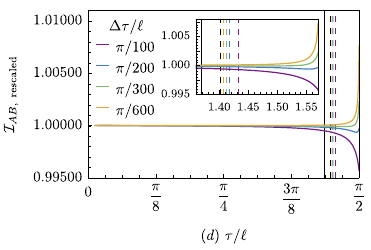}
    \caption{Plots (a), (b), and (c) show the mutual information $\mathcal{I}_{AB}$ between both detectors for $\Delta\tau/\ell=\pi/100$, $\pi/200$, and $\pi/300$ respectively. Observe the formation of the inflection point. Plot (d) gives the rescaled mutual information for $\Delta\tau/\ell=\pi/100$, $\pi/200$, $\pi/300$, and $\pi/600$, such that each function begins at $1$. For all plots, we fix $r_0/r_h=5$, $E\ell=1$, $\zeta=-1$, $M=1$, and $T/\ell=\pi/100+0.001$, and we calculate the image sum from $n=-N$ to $n=+N$ for $N=8$. The solid, vertical line indicates when the support of $\chi_A$ reaches the horizon of the black hole. The first dashed vertical line indicates when the support of $\chi_B$ reaches the horizon, while the second dashed line indicates when the supports of both switching functions are fully contained inside the horizon. In (d), the location of the second dashed line is colour-coded to match each value of $\Delta\tau/\ell$. The right edge of the plot is the time to singularity.
    }
    \label{fig:width1}
\end{figure*}

Fig. \ref{fig:width1}(a)-(c) show the mutual information $\mathcal{I}_{AB}$ between the two detectors as a function of the proper time of detector $A$, given switching function widths $\Delta\tau/\ell=\pi/100$, $\pi/200$, and $\pi/300$ respectively. In each plot, we fix $r_0/r_h=5$, $E\ell=1$, $\zeta=-1$, $M=1$, and $T/\ell=\pi/100+0.001$. In Fig. \ref{fig:width1}(d), we compare the mutual information for $\Delta\tau/\ell=\pi/100$, $\pi/200$, $\pi/300$, and $\pi/600$ on the same plot. The comparison is facilitated by rescaling $\mathcal{I}_{AB}$ such that the first point on each curve takes value 1.

From Fig. \ref{fig:width1}(a)-(c), we observe the formation of an inflection point in $\mathcal{I}_{AB}$ when the width is decreased to roughly $\Delta\tau/\ell=\pi/200$. The behaviour of this inflection point is analogous to that in the case of decreasing black hole mass. The inflection point first appears near the singularity (i.e., the right edge of the plot) and gradually migrates to earlier times in the trajectory. To the right of the inflection point, the mutual information is monotonically increasing. Mutual information grows quickly past the inflection point: the magnitude of the increase in $\mathcal{I}_{AB}$ (for small enough widths) overshadows the magnitude of the decrease leading up to the inflection point. The minimum mutual information attained as a fraction of the mutual information at the first observed point tends to $1$ as $\Delta\tau$ decreases.

One difference between the effect of varying the mass of the black hole versus the effect of varying the width of the detectors' switching function is that the mutual information decreases as $\Delta\tau$ decreases, but there is no analogous scaling as $M$ decreases. Despite the drop in the magnitude of $\mathcal{I}_{AB}$ as $\Delta\tau$ decreases, decreasing $\Delta\tau$ simultaneously facilitates correlation harvesting, as evidenced by the formation of an inflection point. Narrower switching functions uncover correlation by suppressing noise (i.e., $P_A$ and $P_B$) relative to the correlation term. In the case of $\Delta\tau/\ell=\pi/600$ in Fig. \ref{fig:width1}(d), mutual information is monotonically \textit{increasing} from the start of the trajectory.

\begin{figure*}[t]
    \centering
    \includegraphics[width=0.328\linewidth]{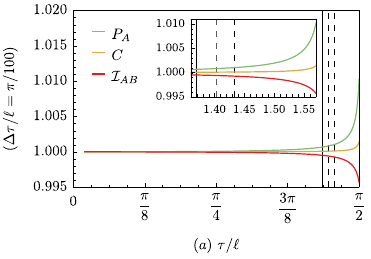}
    \includegraphics[width=0.328\linewidth]{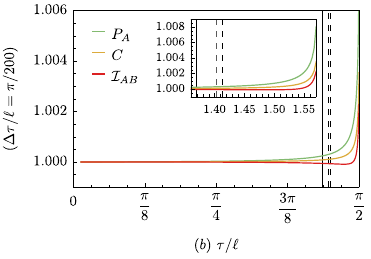}
    \includegraphics[width=0.328\linewidth]{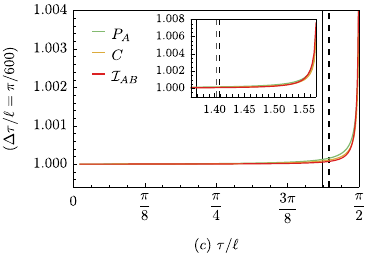}
    \caption{Comparison of transition probability $P_A$, correlation term $C$, and mutual information $\mathcal{I}_{AB}$ for (a) $\Delta\tau/\ell=\pi/100$, (b) $\Delta\tau/\ell=\pi/300$, and (c) $\Delta\tau/\ell=\pi/600$. $P_A$, $C$, and $\mathcal{I}_{AB}$ are rescaled such that each function begins at $1$. Observe that $C$ grows faster relative to $P_A$ as mass decreases, driving the growth of $\mathcal{I}_{AB}$. We fix $r_0/r_h=5$, $E\ell=1$, $\zeta=-1$, $M=1$, and $T/\ell=\pi/100+0.001$, and calculate the image sum from $n=-N$ to $n=+N$ for $N=8$. The solid, vertical line indicates when the support of $\chi_A$ reaches the horizon of the black hole. The dashed, vertical lines indicate when the support of $\chi_B$ reaches the horizon (left line) and when the supports of both switching functions are fully contained inside the horizon (right line). The right edge of the plot is the time to singularity.
    }
    \label{fig:width2}
\end{figure*}

In Fig. \ref{fig:width2}, we compare the transition probability, correlation term, and overall mutual information for switching function widths (a) $\Delta\tau/\ell=\pi/100$, (b) $\Delta\tau/\ell=\pi/300$, and (c) $\Delta\tau/\ell=\pi/600$. In each plot, we fix $r_0/r_h=5$, $E\ell=1$, $\zeta=-1$, $M=1$, and $T/\ell=\pi/100+0.001$. Note again that we only show $P_A$, the transition probability of detector $A$, since $P_B(\tau)=P_A(\tau-T)$. We rescale $P_A$, $C$, and $\mathcal{I}_{AB}$ so that they all begin at $1$.

In Fig. \ref{fig:width2}(a), for the widest switching function, $P_A$ grows more quickly than $C$ as the detectors approach the singularity. As a result, mutual information decreases monotonically. In Fig. \ref{fig:width2}(b)-(c), for narrower switching functions, $C$ increases relative to $P_A$ and becomes comparable to the latter. This growth in $C$ corresponds to the formation of an inflection point and the subsequent monotonically increasing behaviour of the mutual information. If $\Delta\tau$ were decreased further, the growth of $C$ would begin to outpace $P_A$. Overall, the qualitative behaviour of mutual information when decreasing $\Delta\tau$ is analogous to that when decreasing $M$.

\begin{figure*}[t]
    \centering
    \includegraphics[width=0.32\linewidth]{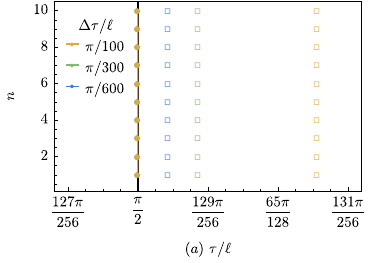}
    \includegraphics[width=0.319\linewidth]{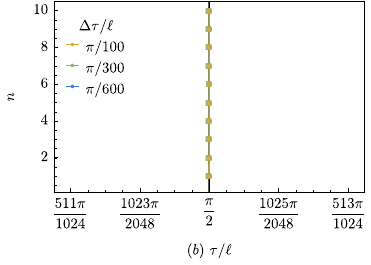}
    \includegraphics[width=0.32\linewidth]{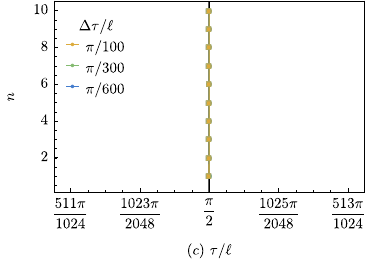}
    \includegraphics[width=0.32\linewidth]{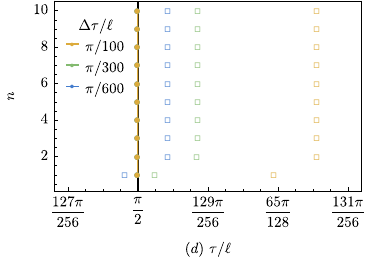}
    \includegraphics[width=0.316\linewidth]{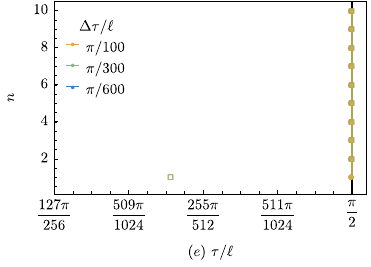}
    \includegraphics[width=0.32\linewidth]{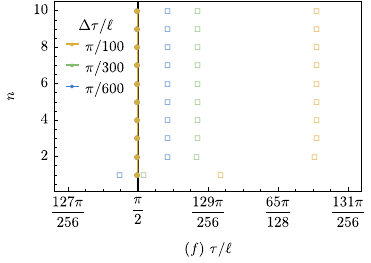}
    \caption{Comparison of the glitches in $C$ (unfilled squares) to the glitches in $P_A$ (filled circles) for $\Delta\tau/\ell=\pi/100$, $\pi/300$, and $\pi/600$. Points having the same colour correspond to the same width $\Delta\tau$. Plots (a), (b), and (c) show the type I, II, and III glitches respectively from the positive term in $C$, while plots (d), (e), and (f) show the type I, II, and III glitches respectively from the negative term in $C$. The $P_A$ glitches are the same in all plots. We fix $r_0/r_h=5$, $M=1$, and $T/\ell=\pi/100+0.001$. The vertical axis indicates the term number $n$ in the image sum from which the glitch arises. The solid vertical line indicates when the black hole singularity is reached.
    }
    \label{fig:width3}
\end{figure*}

Fig. \ref{fig:width3} shows the glitches of the correlation term $C$ relative to the transition probability $P_A$ for $\Delta\tau/\ell=\pi/100$, $\pi/300$, and $\pi/600$. In each plot, we fix $r_0/r_h=5$, $M=1$, and $T/\ell=\pi/100+0.001$. Note again that we only include $P_A$, the transition probability of detector $A$, since $P_B(\tau)=P_A(\tau-T)$. We extend the plot past the time to singularity $\tau/\ell=\pi/2$ in order to show the structure of the glitches.

First, we observe that not all of the types of glitches in $C$ depend on $\Delta\tau$, which can be readily seen from Eq. (\ref{eq:tau11}-\ref{eq:tau23}). However, those that do depend on $\Delta\tau$ move to the left as $\Delta\tau$ decreases. As seen in the previous section with black hole mass, the leftward movement of the glitches in $C$ corresponds to more rapid growth of $C$ along the trajectory, which in turn corresponds to the formation of an inflection point in $\mathcal{I}_{AB}$. Thus, our observations in Fig. \ref{fig:width1}, \ref{fig:width2}, and \ref{fig:width3} are consistent.

In Fig. \ref{fig:width3}(e), in which there is no $\Delta\tau$ dependence in the glitches, the $n=1$ glitch of $C$ always occurs before the first glitch of $P_A$. However, since the position of the glitch does not change, it does not contribute to the dynamics of $\mathcal{I}_{AB}$ as $\Delta\tau$ varies. Thus, we re-emphasize that the absolute position of a glitch does not determine whether an inflection point will definitely be present; it is the relative evolution of the glitches of $C$ with respect to $P_A$ when a parameter is varied that plays the largest role in the formation of an inflection point.

\subsection{Varying the Width and Temporal Separation between Switching Functions} \label{subsec:results-d}

\begin{figure*}[t]
    \centering
    \includegraphics[width=0.49\linewidth]{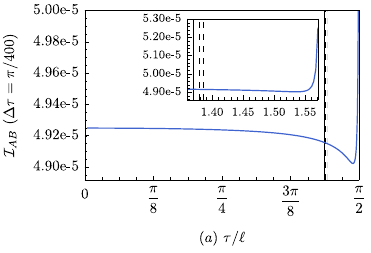}
    \includegraphics[width=0.49\linewidth]{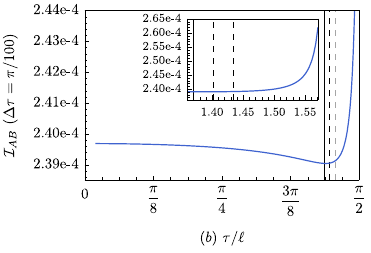}
    \includegraphics[width=0.49\linewidth]{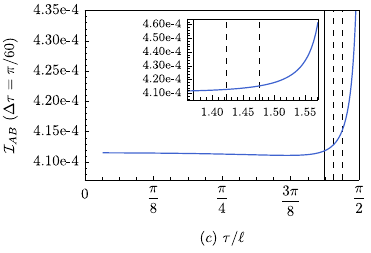}
    \includegraphics[width=0.49\linewidth]{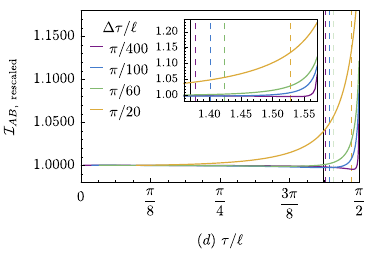}
    \caption{Plots (a), (b), and (c) show the mutual information $\mathcal{I}_{AB}$ between both detectors for $\Delta\tau/\ell=T/\ell-0.001=\pi/400$, $\pi/100$, and $\pi/60$ respectively. Observe the degradation of the inflection point. Plot (d) gives the rescaled mutual information for $\Delta\tau/\ell=T/\ell-0.001=\pi/400$, $\pi/100$, $\pi/60$, and $\pi/20$, such that each function begins at $1$. For all plots, we fix $r_0/r_h=5$, $E\ell=1$, $\zeta=-1$, and $M=0.1$. We calculate the image sum from $n=-N$ to $n=+N$ for $N=22$ when $\Delta\tau/\ell=\pi/400$ and $\Delta\tau/\ell=\pi/100$ and for $N=24$ when $\Delta\tau/\ell=\pi/60$ and $\Delta\tau/\ell=\pi/20$. The solid, vertical line indicates when the support of $\chi_A$ reaches the horizon of the black hole. The  first dashed vertical line indicates when the support of $\chi_B$ reaches the horizon, while the second dashed line indicates when the supports of both switching functions are fully contained inside the horizon. In (d), only the first line (when $\chi_B$ reaches the horizon) is shown, and it is colour-coded to match each value of $\Delta\tau/\ell$. The right edge of the plot is the time to singularity.
    }
    \label{fig:scale1}
\end{figure*}

\begin{figure*}[t]
    \centering
    \includegraphics[width=0.49\linewidth]{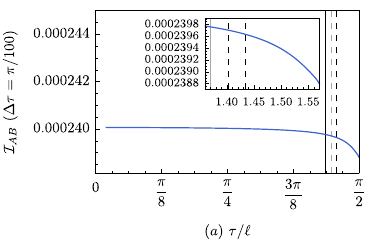}
    \includegraphics[width=0.49\linewidth]{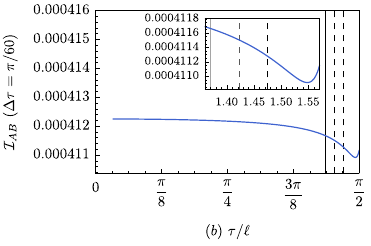}
    \caption{The mutual information $\mathcal{I}_{AB}$ between the two UDW detectors, with $r_0/r_h=5$, $E\ell=1$, $\zeta=-1$, and $M = 0.5$, for (a) $\Delta\tau/\ell=T/\ell-0.001=\pi/100$ and (b) $\Delta\tau/\ell=T/\ell-0.001=\pi/60$. Observe the formation of the inflection point. In both plots, we calculate the image sum from $n=-N$ to $n=+N$ for $N=12$. The solid, vertical line indicates when the support of $\chi_A$ reaches the horizon of the black hole. The dashed, vertical lines indicate when the support of $\chi_B$ reaches the horizon (left line) and when the supports of both switching functions are fully contained inside the horizon (right line). The right edge of the plot is the time to singularity.
    }
    \label{fig:scale2}
\end{figure*}

Fig. \ref{fig:scale1}(a)-(c) show the mutual information $\mathcal{I}_{AB}$ between the two detectors as a function of the proper time of detector $A$, given switching function widths $\Delta\tau/\ell=\pi/400$, $\pi/100$, and $\pi/60$ respectively. At the same time, we change the temporal separation $T$ that $T/\ell=\Delta\tau/\ell+0.001$. In each plot, we fix $r_0/r_h=5$, $E\ell=1$, $\zeta=-1$, and $M=0.1$. In Fig. \ref{fig:scale1}(d), we compare the mutual information for $\Delta\tau/\ell=T/\ell-0.001=\pi/400$, $\pi/100$, $\pi/60$, and $\pi/20$ on the same plot. The comparison is facilitated by rescaling $\mathcal{I}_{AB}$ such that the first point on each curve takes value 1.

As we allow $T$ and $\Delta\tau$ to scale with each other, we observe that the inflection point moves to the left (i.e., to earlier times in the trajectory) as $T$ and $\Delta\tau$ increase. As $\Delta\tau/\ell$ increases from $\pi/400$ to $\pi/20$, the inflection point degrades and eventually disappears, as is the case when $\Delta\tau/\ell=\pi/20$, leaving behind a monotonically increasing mutual information function.

Furthermore, in Fig. \ref{fig:scale2} shows the mutual information $\mathcal{I}_{AB}$ between the two detectors for a different black hole mass, $M=0.5$, given (a) $\Delta\tau/\ell=T/\ell-0.001=\pi/100$ and (b) $\Delta\tau/\ell=T/\ell-0.001=\pi/60$. As before, we fix $r_0/r_h=5$, $E\ell=1$, and $\zeta=-1$. Under these parameters, we observe that an inflection point forms as $T$ and $\Delta\tau$ increase. Overall, the behaviour of mutual information as both $T$ and $\Delta\tau$ increase is analogous to that when $M$ decreases or when $\Delta\tau$ decreases for a fixed $T$: the monotonically decreasing function $\mathcal{I}_{AB}(\tau)$ develops an inflection point that, as one continues to adjust the given parameter, moves to the left and eventually leaves behind a monotonically increasing function.

As with the case of decreasing $\Delta\tau$ for a fixed $T$, which we analyzed in the previous section, the magnitude of $\mathcal{I}_{AB}$ also decreases here as $\Delta\tau$ decreases, even though $T$ decreases as well. That is, the suppression of $\mathcal{I}_{AB}$ due to the narrowing of the switching functions is not fully offset by moving the switching functions closer to each other temporally. Also curiously, we find that it is \textit{increasing} $T$ and $\Delta\tau$ that facilitates the formation of an inflection point. Recall from Fig. \ref{fig:width1} that, for a fixed $T$, the inflection point forms as $\Delta\tau$ decreases. Thus, we might expect that selecting both a smaller $\Delta\tau$ and a smaller temporal separation $T$ would hasten the formation of the inflection point; however, $T$ appears to act in the opposite direction. This curiosity can be better understood when examining the glitches.

\begin{figure*}[t]
    \centering
    \includegraphics[width=0.328\linewidth]{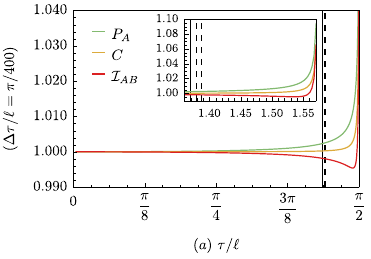}
    \includegraphics[width=0.328\linewidth]{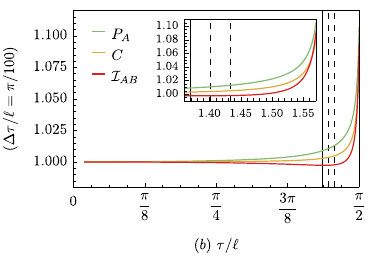}
    \includegraphics[width=0.328\linewidth]{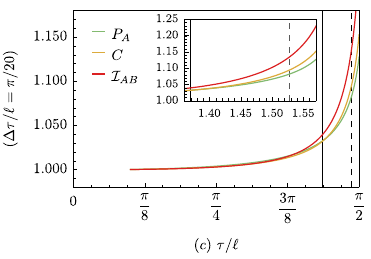}
    \caption{Comparison of transition probability $P_A$, correlation term $C$, and mutual information $\mathcal{I}_{AB}$ for (a) $\Delta\tau/\ell=T/\ell-0.001=\pi/400$, (b) $\Delta\tau/\ell=T/\ell-0.001=\pi/100$, and (c) $\Delta\tau/\ell=T/\ell-0.001=\pi/20$. $P_A$, $C$, and $\mathcal{I}_{AB}$ are rescaled such that each function begins at $1$. Observe that $C$ grows faster relative to $P_A$ as mass decreases, driving the growth of $\mathcal{I}_{AB}$. We fix $r_0/r_h=5$, $E\ell=1$, $\zeta=-1$, and $M=0.1$. The image sum is calculated from $n=-N$ to $n=+N$ for $N=22$ when $\Delta\tau/\ell=\pi/400$ and $\Delta\tau/\ell=\pi/100$ and for $N=24$ when $\Delta\tau/\ell=\pi/20$. The solid, vertical line indicates when the support of $\chi_A$ reaches the horizon of the black hole. The dashed, vertical lines indicate when the support of $\chi_B$ reaches the horizon (left line) and when the supports of both switching functions are fully contained inside the horizon (right line). The right edge of the plot is the time to singularity.
    }
    \label{fig:scale3}
\end{figure*}

In Fig. \ref{fig:scale3}, we compare the transition probability, correlation term, and overall mutual information for switching function widths and temporal separations of (a) $\Delta\tau/\ell=T/\ell-0.001=\pi/400$, (b) $\Delta\tau/\ell=T/\ell-0.001=\pi/100$, and (c) $\Delta\tau/\ell=T/\ell-0.001=\pi/20$. In each plot, we fix $r_0/r_h=5$, $E\ell=1$, $\zeta=-1$, and $M=0.1$. Note again that we only show $P_A$, the transition probability of detector $A$, since $P_B(\tau)=P_A(\tau-T)$. We rescale $P_A$, $C$, and $\mathcal{I}_{AB}$ so that they all begin at $1$.

Phenomenologically, we observe that the growth of the inflection point once again corresponds to the growth of $C$ relative to $P_A$. $C$ grows most slowly relative to $P_A$ in Fig. \ref{fig:scale3}(a), where the inflection point in $\mathcal{I}_{AB}$ is nearest to the singularity, and most quickly in Fig. \ref{fig:scale3}(c), where $\mathcal{I}_{AB}$ is monotonically increasing.

\begin{figure*}[t]
    \centering
    \includegraphics[width=0.32\linewidth]{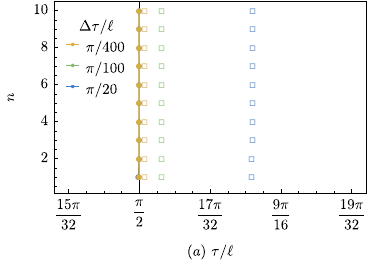}
    \includegraphics[width=0.318\linewidth]{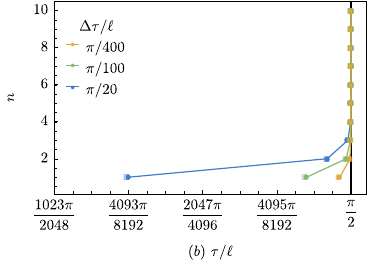}
    \includegraphics[width=0.32\linewidth]{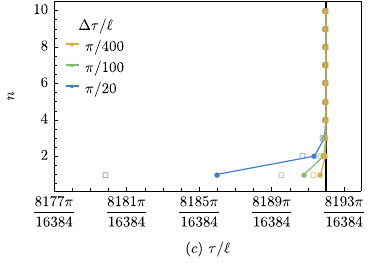}
    \includegraphics[width=0.32\linewidth]{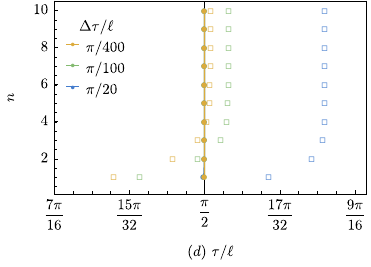}
    \includegraphics[width=0.319\linewidth]{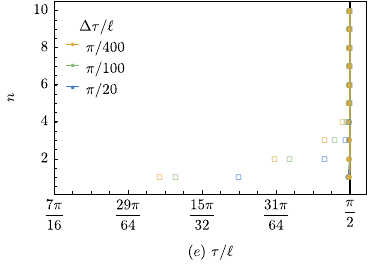}
    \includegraphics[width=0.32\linewidth]{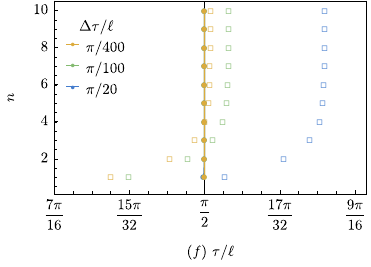}
    \caption{Comparison of the glitches in $C$ (unfilled squares) to the glitches in $P_A$ (filled circles) for $\Delta\tau/\ell=T/\ell-0.001=\pi/400$, $\pi/100$, and $\pi/20$. Points having the same colour correspond to the same values of $\Delta\tau$ and $T$. Plots (a), (b), and (c) show the type I, II, and III glitches respectively from the positive term in $C$, while plots (d), (e), and (f) show the type I, II, and III glitches respectively from the negative term in $C$. The $P_A$ glitches are the same in all plots. We fix $r_0/r_h=5$ and $M=0.1$. The vertical axis indicates the term number $n$ in the image sum from which the glitch arises. The solid vertical line indicates when the black hole singularity is reached.
    }
    \label{fig:scale4}
\end{figure*}

Fig. \ref{fig:scale4} shows the glitches of the correlation term $C$ relative to the transition probability $P_A$ for $\Delta\tau/\ell=T/\ell-0.001=\pi/400$, $\pi/100$, and $\pi/20$. In each plot, we fix $r_0/r_h=5$ and $M=0.1$. Note again that we only include $P_A$, the transition probability of detector $A$, since $P_B(\tau)=P_A(\tau-T)$. We extend the plot past the time to singularity $\tau/\ell=\pi/2$ in order to show the structure of the glitches.

In Fig. \ref{fig:scale4}(a), (d), (e), and (f), the glitches in $C$ move to the right as $\Delta\tau$ and $T$ increase. Ignoring the fact that $T$ is also varying here, this behaviour is similar to what was observed in Fig. \ref{fig:width3}, where larger $\Delta\tau$ corresponded to glitches further to the right. However, in Fig. \ref{fig:scale4}(b)-(c), the glitches in $C$ move to the \textit{left} as $\Delta\tau$ and $T$ increase--something that was not seen in Fig. \ref{fig:width3} when only $\Delta\tau$ varied. Moreover, the leftward movement of the $C$-glitches in Fig. \ref{fig:scale4}(b)-(c) keeps pace with or even outpaces that of the $P_A$-glitches. The result is that an inflection point is induced as both $\Delta\tau$ and $T$ are varied. This occurs despite Fig. \ref{fig:scale4}(a), (d), (e), and (f) showing the opposite trend---recall the comment from section \ref{subsec:results-b} that only one set of $C$-glitches, growing leftward at sufficient pace, is needed to generate the required growth in $C$ itself. We find, as with the other parameters we had investigated, that the glitches here are also consistent with the observations made in the other plots (that is, Fig. \ref{fig:scale1}-\ref{fig:scale3}) in this section.

\subsection{Varying Detector Energy Gap}  \label{subsec:results-e}

\begin{figure*}[t]
    \centering
    \includegraphics[width=0.49\linewidth]{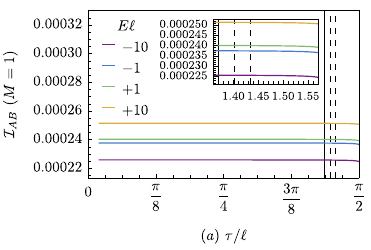}
    \includegraphics[width=0.49\linewidth]{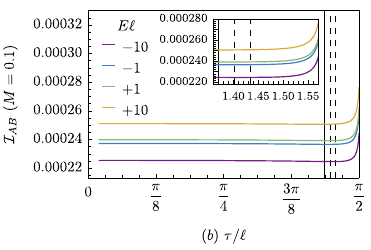}
    \caption{The mutual information $\mathcal{I}_{AB}$ between the two UDW detectors varying the energy gap $E\ell$, with $r_0/r_h=5$, $\zeta=-1$, $\Delta\tau/\ell=\pi/100$, and $T/\ell=\Delta\tau/\ell+0.001$, for (a) $M = 1$ and (b) $M = 0.1$. We calculate the image sum from $n=-N$ to $n=+N$ for $N=8$ and $22$ respectively. The solid, vertical line indicates when the support of $\chi_A$ reaches the horizon of the black hole. The dashed, vertical lines indicate when the support of $\chi_B$ reaches the horizon (left line) and when the supports of both switching functions are fully contained inside the horizon (right line). The right edge of the plot is the time to singularity.
    }
    \label{fig:energies}
\end{figure*}

Fig. \ref{fig:energies} shows the mutual information $\mathcal{I}_{AB}$ between the two detectors as a function of the proper time of detector $A$ for four detector energy gaps, $E\ell=-10$, $-1$, $+1$, and $+10$, given black hole mass of (a) $M=1$ and (b) $M=0.1$. In both plots, we fix $\zeta=-1$, $\Delta\tau/\ell=\pi/100$, and $T/\ell=\Delta\tau/\ell+0.001$.

Changing the energy gap of the detectors results in a vertical shift in mutual information, with larger energy gaps corresponding to more harvested correlation. No additional structure in the mutual information function arises because the locations of the glitches do not depend on the energy gap, thus we do not observe changes in the relative growth of $P_D$ and $C$. These results justify the fact that we have fixed $E\ell=+1$ in all of the previous sections.

\section{Conclusion}\label{sec:conclusion}

Perhaps the most salient point of our investigation is that black holes---even the simplest kind having constant curvature---significantly affect the correlation properties of quantum fields in the vacuum state.
Our results are commensurate with previous work ~\cite{MutualInfoStatic2022} stating that detectors can harvest mutual information up to the edge of the horizon; we have shown that freely falling detectors can harvest it across and within the horizon.

Our study was motivated by the growing literature regarding entanglement and correlation harvesting in the vicinity of a black hole and regarding the observables of detectors that fall behind a black hole horizon. We noticed, however, a scarcity of literature that considers both of these problems simultaneously. Following a recent study on the entanglement harvesting of UDW detectors behind the horizon of a (1+1)-dimensional Schwarzschild spacetime ~\cite{Gallock-Yoshimura:2021yok}, we thus computed the mutual information for two UDW detectors falling into the horizon of a (2+1)-dimensional BTZ black hole. Unlike the (1+1)-dimensional Schwarzschild black hole, the BTZ black hole satisfies Einstein's equations.

We numerically calculated the mutual information between two freely falling UDW detectors coupled to a massless conformal scalar field in a non-rotating (2+1)-dimensional BTZ black hole. We investigated the effects of varying the boundary condition of the field, the black hole mass, the width of the detectors' switching functions, the temporal separation between the switching functions (along with the width), and the energy gap of the detectors. The mutual information $\mathcal{I}_{AB}$ was computed to lowest non-vanishing order in the coupling constant $\lambda$. We expressed the Wightman function of the BTZ black hole as an image sum of Wightman functions in AdS$_3$, and then computed $P_A$, $P_B$, and $C$ as a sum of the contributions arising from each image, where $P_{A,B}$ are the transition probabilities of the detectors and $C$ is a correlation term. This process was repeated at multiple points along the trajectory of the detectors in order to generate a mutual information curve for each simulation.

The mutual information is approximately constant when the detectors are far away from the black hole, since the spacetime there is like AdS. However, as the detectors fall, the mutual information begins to deviate from its constant value. The three cases that we observed were: (i) mutual information decreasing monotonically as $r\rightarrow 0$; (ii) mutual information decreasing, attaining an inflection point, and then increasing as $r\rightarrow 0$; and (iii) mutual information increasing monotonically as $r\rightarrow 0$. These three cases are not isolated; the mutual information satisfying one case can be gradually transformed to satisfy the next case by continuously varying certain parameters in the problem.

We observed that the mass $M$ of the black hole, the width $\Delta\tau$ of the detector switching functions, and the temporal separation $T$ between the switching functions had the largest effect on mutual information. Specifically, decreasing $M$, decreasing $\Delta\tau$, or increasing both $\Delta\tau$ and $T$ proportionally resulted in the mutual information changing from case (i) to case (ii) and then to case (iii). On the other hand, the energy gap $E$ of the detectors had no impact on mutual information except to introduce a vertical shift to the values of $\mathcal{I}_{AB}$ along the trajectory. The field boundary condition $\zeta$ also had a modest impact on the mutual information between the detectors. Specifically, we observed the inflection point to form most readily when $\zeta=-1$, followed by $\zeta=0$ and then $\zeta=1$.

One limitation of our study is that we have considered a (2+1)-dimensional spacetime with constant curvature, whereas physical black holes are (3+1)-dimensional with varying spacetime curvature. The main reason for working in a lower-dimensional spacetime (particularly, the BTZ spacetime) is that the Wightman function is tractable. Nonetheless, we expect the results of our study to provide insights on the problem of correlation harvesting behind the horizon that may be applied to higher-dimensional spacetimes. Another constraint in our study is that we have computed the mutual information, which is the sum of quantum and classical correlations, between the detectors, rather than pure entanglement harvesting. We compute mutual information because the dynamical problem of harvesting entanglement as detectors cross the event horizon is substantially more difficult. A logical follow-up would be to calculate the entanglement harvesting in a similar setup between two detectors crossing an event horizon. Another immediate undertaking would be to investigate other trajectories for the detectors.

Other future avenues of research include examining the entanglement harvesting between detectors that fall behind the horizon of a rotating black hole, or in a black hole spacetime where the curvature is not constant. The rotating problem is particularly interesting because rotation has been observed to amplify entanglement harvesting ~\cite{Robbins:2020jca}. Additionally, the rotating BTZ black hole possesses a Wightman function that is asymmetric in the image sum as $n\rightarrow -n$ and also possesses a more complex glitch structure ~\cite{Sijia2024}. The latter point is salient since we have found, in this study, that glitches are correlated with changes in mutual information along the detectors' trajectory.

\section*{Acknowledgments}

This work was supported in part by the Natural Sciences and Engineering Research Council of Canada. MRPR gratefully acknowledges the support provided by the Mike and Ophelia Lazaridis Graduate Fellowship.

\newpage

\bibliography{ref}

\end{document}

%% file: preamble.tex
\pdfoutput=1 
\usepackage[english]{babel}
\usepackage[T1]{fontenc} 
\usepackage{amsmath,amssymb,amsfonts,amsthm}
\usepackage{graphicx}
\usepackage[usenames,dvipsnames]{color}
\usepackage{enumitem}
\usepackage{verbatim}
\usepackage[percent]{overpic}
\usepackage{rotating}
\usepackage[colorlinks=true,allcolors=black]{hyperref}
\usepackage{array}
\usepackage{mathtools}
\usepackage{lmodern}
\usepackage[normalem]{ulem}
\usepackage{soul} 
\usepackage{braket}
\usepackage{tensor}
\usepackage{dsfont}
\usepackage{bm}
\usepackage{tikz}

\usepackage{mathrsfs}

\usetikzlibrary{decorations.pathmorphing}
\usepackage{CJKutf8}

\DeclareMathOperator\arctanh{arctanh}

\definecolor{patriarch}{rgb}{0.5, 0.0, 0.5}
\definecolor{darkraspberry}{rgb}{0.53, 0.15, 0.34}
\definecolor{brinkpink}{rgb}{0.98, 0.38, 0.5}
\definecolor{skobeloff}{rgb}{0.0, 0.48, 0.45}
\definecolor{mypink}{RGB}{226,68,130}
\definecolor{darkpastelgreen}{rgb}{0.01, 0.75, 0.24}
\definecolor{pigmentgreen}{rgb}{0.0, 0.65, 0.31}

\usetikzlibrary{positioning}
\usetikzlibrary{calc,through,backgrounds}

\usepackage{bbm}

\newcommand{\R}{\mathbb{R}}
\newcommand{\dd}{\text{d}}

\newcommand{\sx}{\mathsf{x}}

\newcommand{\vp}{\varphi}
\newcommand{\ii}{\mathsf{i}}

%% file: title_and_authors.tex
\title{Harvesting Information Across the Horizon}

\author{Sijia Wang}
\email{s676wang@uwaterloo.ca}
\affiliation{Department of Physics and Astronomy, University of Waterloo, Waterloo, Ontario, N2L 3G1, Canada}
\affiliation{Waterloo Centre for Astrophysics, University of Waterloo, Waterloo, Ontario, N2L 3G1, Canada}

\author{María Rosa Preciado-Rivas}
\email{mrpreciadorivas@uwaterloo.ca} 
\affiliation{Waterloo Centre for Astrophysics, University of Waterloo, Waterloo, Ontario, N2L 3G1, Canada}
\affiliation{Department of Applied Mathematics, University of Waterloo, Waterloo, Ontario, N2L 3G1, Canada}
\affiliation{Institute for Quantum Computing, University of Waterloo, Waterloo, Ontario, N2L 3G1, Canada}

\author{Robert B. Mann}
\email{rbmann@uwaterloo.ca}
\affiliation{Department of Physics and Astronomy, University of Waterloo, Waterloo, Ontario, N2L 3G1, Canada}
\affiliation{Institute for Quantum Computing, University of Waterloo, Waterloo, Ontario, N2L 3G1, Canada}
\affiliation{Perimeter Institute for Theoretical Physics,  Waterloo, Ontario, N2L 2Y5, Canada}

